\documentclass[fleqn,twoside]{article}
\usepackage{espcrc2}
\usepackage{float}

\usepackage{graphicx}

\usepackage[figuresright]{rotating}

\newcommand{\AmS}{{\protect\the\textfont2
  A\kern-.1667em\lower.5ex\hbox{M}\kern-.125emS}}

\def\lesssim{\mathrel{\hbox{\rlap{\hbox{\lower4pt\hbox{$\sim$}}}\hbox{$<$}}}}
\def\gtrsim{\mathrel{\hbox{\rlap{\hbox{\lower4pt\hbox{$\sim$}}}\hbox{$>$}}}}

\title{Highest Energy Neutrino Showers in EUSO}
\author{D.Fargion \address[Uniroma1]
        {Physics department, Universita' degli studi "La Sapienza", \\
         Piazzale Aldo Moro 5, -  00185 Roma, Italy}%
        \address[INFN]
        {INFN Roma, Istituto Nazionale di Fisica Nucleare,
        Italy}\\
}
\begin{document}
\begin{abstract}
EUSO experiment, while monitoring the downward Earth atmosphere
layers, may observe among common Ultra High Energy Cosmic Rays,
UHECR, also High Energy Neutrino-Induced Showers either blazing
upward to the detectors at  high ($\sim$ PeVs) energies or at much
higher GZK, $\sim E_{\nu}\geq 10^{19}$ eV energies, showering
horizontally in air or vertically downward.  A small fraction of
these upward, horizontal and vertical Shower maybe originated by
direct astrophysical UHE neutrino interacting on terrestrial air
layers itself; however the dominant UHE neutrino signal are
Upward and Horizontal Tau Air-Showers, UPTAUS and HORTAUs (or
Earth skimming $\nu$), born within widest Earth Crust Crown (Sea
or Rock) Areas, by UHE $\nu_{\tau} + Nuclei$ $\rightarrow \tau$
interactions, respectively at PeVs and GZK energies: their rate
and signatures are shown in a neutrino fluence map for EUSO
thresholds versus other UHE air interacting neutrino signals and
backgrounds. The effective target Masses originating HORTAUs seen
by EUSO may exceed (on sea) a wide and huge ring volume $\simeq
5130$ $km^3$. The consequent HORTAUS event rate (even at $10\%$
EUSO duty cycle lifetime) may deeply test the expected Z-Burst
models by at least a hundred of yearly events. Even rarest but
inescapable GZK neutrinos (secondary of photopion production of
observed cosmic UHECR) might be discovered in a few (or a tens)
horizontal shower events; in this view an extension of EUSO
detectability up to $\sim E_{\nu}\geq 10^{19}$eV threshold is to
be preferred. A wider collecting EUSO telescope (3m diameter)
might be considered.

\end{abstract}
\maketitle

-------------------------------------------------

\section*{Introduction: EUSO and GZK $\nu$}

The very possible discover of an UHECR astronomy, the solution of
the GZK paradox, the very urgent rise of an UHE neutrino
astronomy are among the main goals of EUSO project. This advanced
experiment in a very near future will encompass AGASA-HIRES and
AUGER and observe for highest cosmic ray showers on Earth
Atmosphere recording their tracks from International Space
Station by Telescope facing dawn-ward the Earth. Most of the
scientific community is puzzled by the many mysteries of UHECRs:
their origination because of their apparent isotropy, is probably
extragalactic. However the UHECR events are not clustered to any
nearby AGN, QSRs or Known GRBs within the narrow (10-30 Mpc
radius) volume defined by the cosmic 2.75 $K^{o}$ proton drag
viscosity (the so called GZK cut-off
\cite{Greisen:1966jv}\cite{Zatsepin:1966jv}). The recent doublets
and triplets clustering found by AGASA seem to favor compact
object (as AGN) over more exotic topological relic models, mostly
fine tuned in mass (GUT, Planck one) and time decay rate to fit
all the observed spectra. However the missing AGN within a GZK
volume is wondering. A possible remarkable correlation recently
shows that most of the UHECR event cluster point toward BL Lac
sources \cite{Gorbunov Tinyakov Tkachev Troitsky}. This
correlation favors a cosmic origination for UHECRs, well above
the near GZK volume. In this frame a relic neutrino mass
\cite{Dolgov2002}, \cite{Raffelt2002}  $m_{\nu} \simeq 0.4$ eV or
($m_{\nu} \simeq 0.1 \div 5$ eV) may solve the GZK paradox
\cite{Fargion Salis 1997} , \cite{Fargion Mele Salis
1999},\cite{Weiler 1999},\cite{Yoshida et all 1998},\cite{Fargion
et all. 2001b},\cite{Fodor Katz Ringwald 2002} overcoming the
proton opacity being ZeV UHE neutrinos transparent (even from
cosmic edges to cosmic photon Black Body drag) while interacting
in resonance with relic neutrinos masses in dark halos (Z-burst or
Z-WW showering models). These light neutrino masses do not solve
the galactic or cosmic dark matter problem but it is well
consistent with old and  recent solar neutrino oscillation
evidences \cite{Gallex92},\cite{Fukuda:1998mi},\cite{SNO2002} and
most recent claims by KamLAND \cite{Kamland2002} of anti-neutrino
disappearance  (all in agreement within a Large Mixing Angle
neutrino model and $\triangle {m_{\nu}}^2 \sim 7 \cdot
10^{-5}{eV}^2$) as well as these light masses are in agreement
with atmospheric neutrino mass splitting ($\triangle m_{\nu}
\simeq 0.07$ eV) and in fine tune with more recent neutrino
double beta decay experiment mass claim $m_{\nu} \simeq 0.4$ eV
\cite{Klapdor-Kleingrothaus:2002ke}. In this Z-WW Showering for
light neutrino mass models large fluxes of UHE $\nu$ are
necessary,\cite{Fargion Mele Salis 1999},\cite{Yoshida et all
1998}\cite{Fargion et all. 2001b}, \cite{Fodor Katz Ringwald
2002},\cite{Kalashev:2002kx} or higher than usual gray-body
spectra of target relic neutrino or better clustering are needed
\cite{Fargion et all. 2001b}\cite{Singh-Ma}: indeed a heaviest
neutrino mass  $m_{\nu} \simeq 1.2-2.2$ eV while still being
compatible with known bounds, might better gravitationally cluster
leading to denser dark local-galactic halos and lower neutrino
fluxes\cite{Fargion et all. 2001b}\cite{Singh-Ma}. It should
remarked that in this frame the main processes leading to UHECR
above GZK are mainly the WW-ZZ and the t-channel interactions
\cite{Fargion Mele Salis 1999},\cite{Fargion et all. 2001b}.
These expected UHE neutrino fluxes might and must  be
experienced  in complementary and independent tests.
\section{ UHE $\nu$ Astronomy by the $\tau$ Showering }
While longest ${\mu}$ tracks in $km^3$ underground detector have
been, in last three decades, the main searched UHE neutrino
signal, Tau Air-showers by UHE neutrinos generated in Mountain
Chains or within Earth skin crust at Pevs up to GZK ($>10^{19}$
eV) energies have been recently proved to be a  new powerful
amplifier in Neutrino Astronomy \cite{Fargion et all 1999},
\cite{Fargion 2000-2002},\cite{Bertou et all 2002},\cite{Hou
Huang 2002},\cite{Feng et al 2002}. This new Neutrino $\tau$
detector will be (at least) complementary to present and future,
lower energy, $\nu$ underground  $km^3$ telescope projects (from
AMANDA,Baikal, ANTARES, NESTOR, NEMO, IceCube). In particular
Horizontal Tau Air shower may be naturally originated by UHE
$\nu_{\tau}$ at GZK energies crossing the thin Earth Crust at the
Horizon showering far and high in the atmosphere \cite{Fargion
2000-2002},\cite{Fargion2001a}, \cite{Fargion2001b},\cite{Bertou
et all 2002},\cite{Feng et al 2002}. UHE $\nu_{\tau}$ are
abundantly produced by flavour oscillation and mixing from muon
(or electron) neutrinos, because of the large galactic and cosmic
distances respect to the neutrino  oscillation ones (for already
known neutrino mass splitting). Therefore EUSO may observe many
of the above behaviours and it may constrains among models and
fluxes and it may also answer open standing questions. I will
briefly enlist, in this first preliminary presentation, the main
different signatures and rates of UHECR versus UHE $\nu$ shower
observable by EUSO at 10\% duty cycle time within a 3 year record
period, offering a first estimate of their signals. Part of the
results on UHECR are probably  well known, even here it is
re-estimated. Part of the results, regarding the UPTAUs and
HORTAUs, are new and they rule the UHE $\nu$ Astronomy in EUSO.
\begin{figure}\centering\includegraphics[width=8cm]{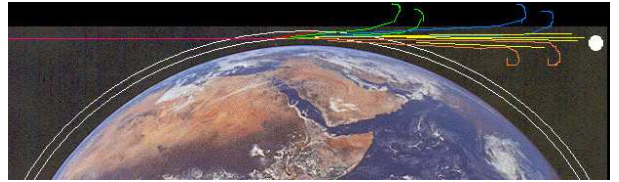}
\vspace{-1.5cm} \caption {A very schematic Horizontal High
Altitude Shower (HIAS); its fan-like imprint is due to
geo-magnetic bending of charged particles at high quota ($\sim 44
km$). The Shower may  point to an satellite as old gamma
GRO-BATSE detectors or very recent Beppo-Sax,Integral, HETE,
Chandra or future Agile and Swift ones. \cite{Fargion
2000-2002},\cite{Fargion2001a},\cite{Fargion2001b}. The HIAS
Showers is open and forked in  five (or three or at least two
main component): ($e^+,e^-,\mu^+,\mu^-, \gamma $, or just
positive-negative); these  multi-finger tails may be seen as split
tails  by EUSO. } \label{fig:fig1} \vspace{-0.3cm}
\end{figure}
\begin{figure}
\vspace{- 0.2cm}
\centering\includegraphics[width=8cm]{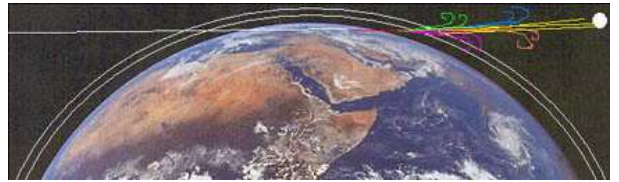}
\vspace{-1cm} \caption { As above  Horizontal Upward Tau
Air-Shower (HORTAUS) originated by UHE neutrino skimming the
Earth: fan-like jets due to geo-magnetic bending  shower at high
quota ($\sim 23-40 km$): they may be pointing to an orbital
satellite detector \cite{Fargion 2000-2002}, \cite{Fargion2001a},
\cite{Fargion2001b}. The Shower tails may be also observable by
EUSO just above it.} \label{fig:fig2}
\end{figure}
\begin{figure}
\vspace{-0.5cm}
\centering\includegraphics[width=8cm]{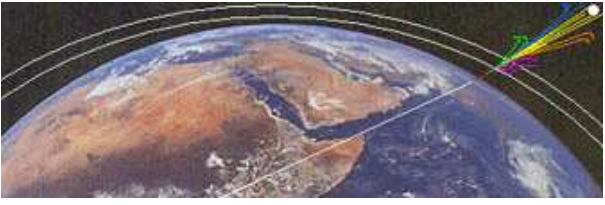}
\caption {A very schematic Upward Tau Air-Shower (UPTAUs)  and its
open fan-like jets due to geo-magnetic bending at high quota
($\sim 20-30 km$). The gamma Shower may be pointing to an orbital
detector \cite{Fargion 2000-2002}, \cite{Fargion2001a},
\cite{Fargion2001b}. Its
 vertical Shower tail may be spread by
geo-magnetic field into a thin eight-shape beam observable  by
EUSO  as a small blazing oval (few dot-pixels) aligned orthogonal
to the local magnetic field .} \label{fig:fig3}
 \vspace{-0.9cm}
\end{figure}

\section{Upward UHE $\nu$  Showering in Air }
Let us first consider the last kind of Upward $\tau$ signals due
to their interaction in Air or in Earth Crust. The Earth opacity
will filter mainly  $10^{14}-10^{15}$eV upward events \cite{Gandhi
et al 1998},\cite{Halzen1998},\cite{Becattini Bottai
2001},\cite{Dutta et al.2001},\cite{Fargion 2000-2002}; therefore
only the direct $\nu$ shower in air or the UPTAUs around PeVs
will be able to flash toward EUSO in a narrow beam ($2.5 \cdot
10^{-5}$ solid angle) jet blazing apparently at
$10^{19}-10^{20}$eV energy. The shower will be opened in a fan
like shape and it will emerge from the Earth atmosphere spread as
a triplet or multi-dot signal aligned orthogonal to local
terrestrial magnetic field. This signature will be easily
revealed. However the effective observed air mass by EUSO is not
$\ 10\%$ (because duty cycle) of the inspected air volume $\sim
150 km^3$, but because of the narrow blazing shower cone it
corresponds to only to $3.72\cdot 10^{-3}$ $km^3$. The target
volume  increases  for upward neutrino Tau interacting vertically
in Earth Crust in last matter layer (either rock or water),
making upward relativistic $\simeq PeVs$ $\tau$ whose decay in
air born finally an UPTAUs; in this case the effective target
mass is  (for water or rock) respectively $5.5\cdot
10^{-2}$$km^3$ or $1.5 \cdot10^{-1}$ $km^3$. These volume are not
extreme. The consequent foreseen thresholds are summirized for
$3$ EUSO years of data recording in Figure $4$. The UPTAUs signal
is nearly $15$ times larger than the Air-Induced $\nu$ Shower. A
more detailed analysis may show an additional factor three (due
to the neutrino flavours) in favor of Air-Induced Showers, but
the more transparent role of PeV multi-generating upward
$\nu_{\tau}$ while crossing the Earth, makes the result
summirized in figure. The much wider acceptance of BATSE respect
EUSO and the consequent better threshold (in BATSE) is due to the
wider angle view of the gamma detector, the absence of any
suppression factor as in EUSO duty cycle, as well as the $10$
(for BATSE) over $3$ (for EUSO) years assumed of record
life-time. Any minimal neutrino  fluence $\Phi_{\nu_{\tau}}$ of
PeVs neutrino
 $$ \Phi_{\nu_{\tau}}\geq 10^2 eV cm^{-2} s^{-1}$$ might be detectable by EUSO.
\begin{figure}\centering\includegraphics[width=8cm]{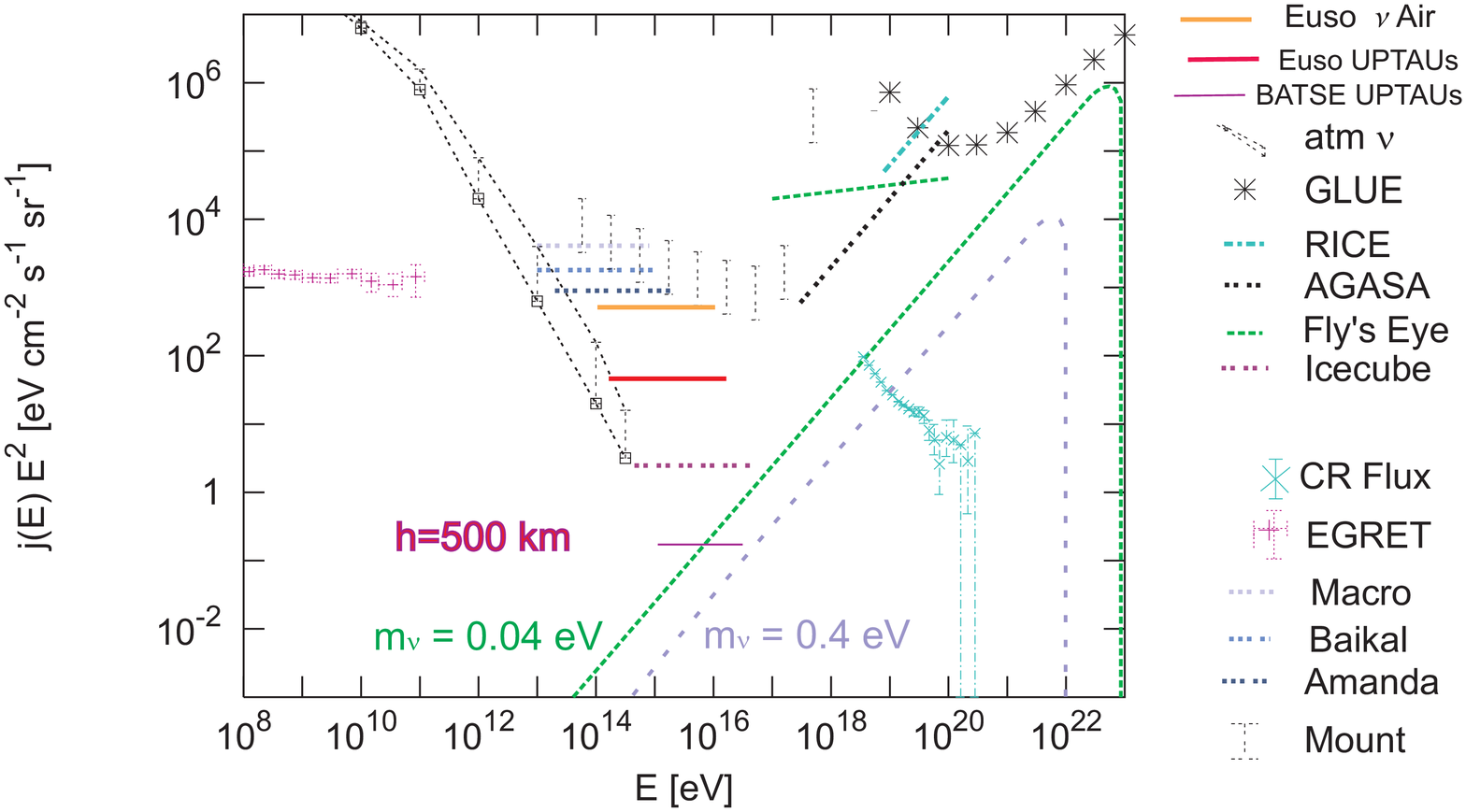}
\caption {Upward Neutrino Air-Shower and Upward Tau Air-shower,
UPTAUs, Gamma and Cosmic Rays Fluence Thresholds and bounds in
different energy windows for different past and future detectors.
The UPTAUs threshold for EUSO has been estimated for a three year
experiment lifetime. BATSE recording limit is also shown from
height $h = 500km$ and for ten year record. Competitive experiment
are also shown as well as the Z-Shower expected spectra in light
neutrino mass values $m_{\nu} = 0.4, 0.04$ eV. \cite{Fargion
2000-2002}, \cite{Fargion2001a},\cite{Kalashev:2002kx},
\cite{Fargion et all. 2001b},\cite{Fargion 2002d}.}
\end{figure}

\section{Downward and Horizontal UHECRs }
Let us now briefly reconsider the nature of common Ultra High
Cosmic Rays (UHECR) showers. Their  rate  will offer a useful test
for any additional UHE neutrino signals.
 Let us assume for sake of simplicity a characteristic opening angle of
EUSO telescope of $30^o$ and a nominal satellite  height of $400$
km, leading to an approximate atmosphere area under inspection of
EUSO $\sim 1.5 \cdot 10^5 km^2$. Let us discuss the UHECR shower:
It has been estimated (and it is easy to verify)  a $\sim
2\cdot10^{3}$ event/year rate above $3\cdot10^{19}$ eV. Among
these "GZK" UHECR (either proton, nuclei or $\gamma$) nearly
$7.45\%\approx 150$ event/year will shower in Air Horizontally
with no Cherenkov hit on the ground. The critical angle $6.7^o$
corresponding to $7.45\%$ of all the events, is derived from
first interacting quota (here assumed for Horizontal Hadronic
Shower near $44$ km following \cite{Fargion
2000-2002},\cite{Fargion2001a},\cite{Fargion2001b}): Indeed the
corresponding horizontal edge critical angle $\theta_{h}$ $=$
$6.7^o$ below the horizons ($\pi{/2}$) is found (for an
interacting height h near $44$ km):$ {\theta_{h} }={\arccos
{\frac {R_{\oplus}}{( R_{\oplus} + h_1)}}}\simeq 1^o \sqrt{\frac
{h_{1}}{km}} $. These Horizontal High Altitude Showers
(HIAS)\cite{Fargion2001a},\cite{Fargion2001b}, will be able to
define a new peculiar showering, mostly very long (hundred kms)
and bent and forked (by few or several degrees) by local
geo-magnetic fields. The total UHECR above $3\cdot10^{19}$ eV
will be $\sim 6000$ UHECR and $\sim 450$ Horizontal Shower within
3 years; these latter horizontal signals are relevant because
they may mimic Horizontal  induced $\nu$ Air-Shower, but mainly
at high quota ($\geq 30-40 km$) and down-ward. On the contrary UHE
neutrino tau showering, HORTAUs, to be discussed later, are also
at high quota ($\geq 23 km$), but  upward-horizontal. Their
outcoming angle will be ($\geq 0.2^o-3^o$) upward. Therefore a
good angular ($\leq 0.2-0.1 ^o$) resolution to distinguish
between the two signal will be a key discriminator. While
Horizontal UHECR are an important piece of evidence in the UHECR
calibration and its GZK study , at the same time they are a
severe back-ground noise competitive with Horizontal-Vertical GZK
Neutrino Showers originated in Air, to be discussed below. However
Horizontal-downward UHECR are not confused with upward Horizontal
HORTAUs by UHE neutrinos to be summirized in last section. Note
that Air-Induced Horizontal UHE neutrino as well as all down-ward
Air-Induced UHE $\nu$ will shower mainly at lower altitudes
($\leq 10 km$) ; however they are respectively only a small
($\leq 2\% $, $\leq 8\%$) fraction than HORTAUs showers to be
discussed in the following. An additional factor $3$ due to their
three flavour over $\tau$ unique one may lead to respectively
($\leq 6\% $, $\leq 24\% $) of all HORTAUs events: a contribute
ratio that may be in principle an useful test to study the
balanced neutrino flavour mixing.


\begin{figure}\centering\includegraphics[width=8cm]{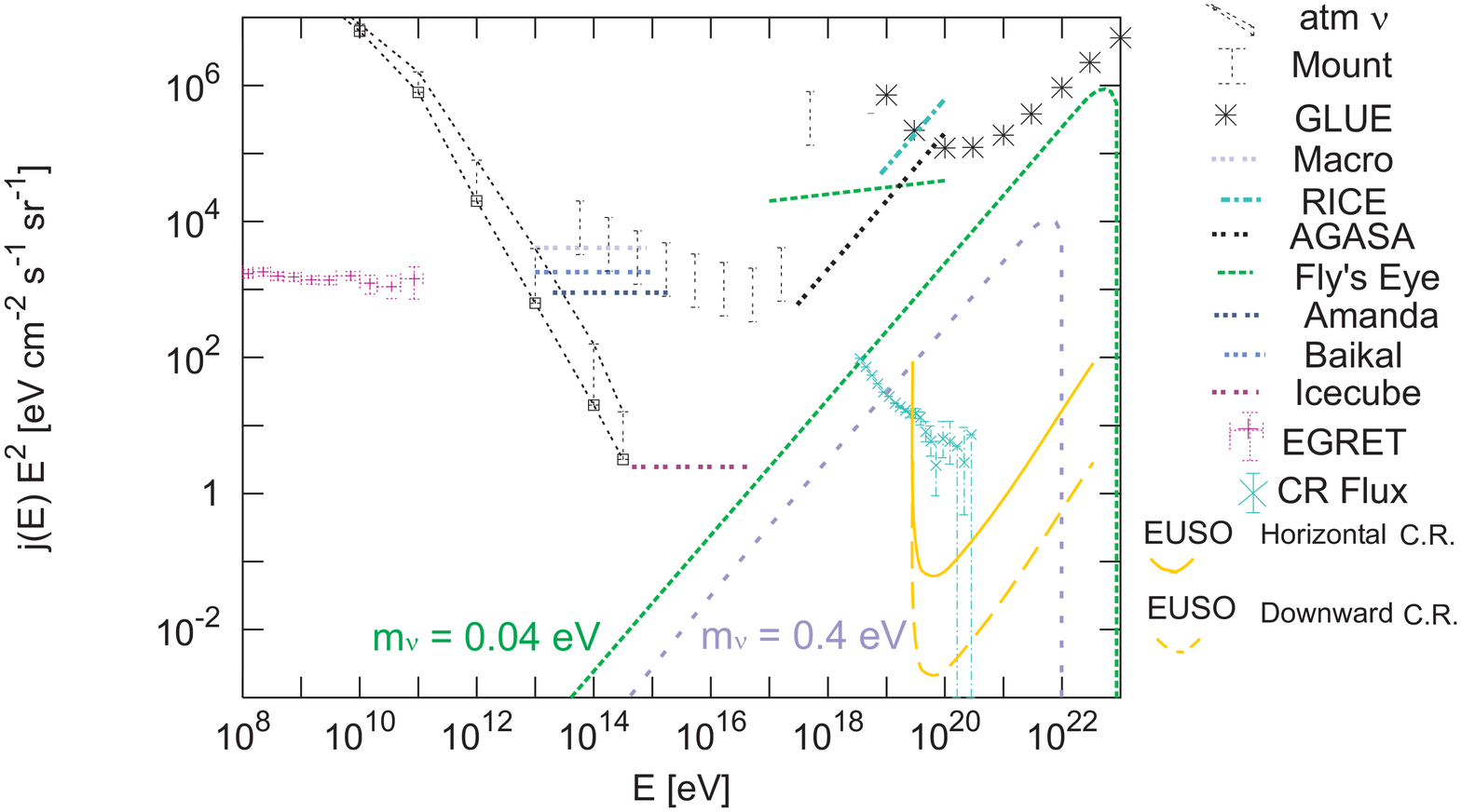}
\caption {Neutrino, Gamma and Cosmic Rays Fluence Thresholds and
bounds in different  energy windows. The Cosmic Rays Fluence
threshold for EUSO has been estimated  for a three year
experiment lifetime. The paraboloid bound shape threshold may
differ upon the EUSO optics and acceptance. Competitive
experiment are also shown as well as the Z-Shower expected
spectra in light mass values. \cite{Fargion 2000-2002},
\cite{Fargion2001a},\cite{Kalashev:2002kx}, \cite{Fargion et all.
2001b},\cite{Fargion 2002d}.} \label{fig:fig2}
\end{figure}


\section{ Air Induced UHE $\nu$   Shower }
 UHE $\nu$ may hit an air nuclei and shower
vertically or horizontally or more rarely nearly up-ward: its
trace maybe observable by EUSO preferentially in inclined or
horizontal case.  Indeed  Vertical Down-ward  ($\theta \leq
60^o$) neutrino induced Air Shower  occur mainly
 at lowest quota and they will only partially shower their UHE $\nu$ energy
 because of the small slant depth ($\leq 10^3 g cm^{-2}$) in most vertical down-ward UHE $\nu$ shower.
 The observed  EUSO air mass ($1500 km^3$, corresponding to a $\sim 150$ $km^3$ for $10\%$ EUSO record time)
  is only ideally the UHE neutrino calorimeter. Indeed  inclined ($\sim{\theta\geq 60^o }$) and horizontal
 Air-Showers ($\sim{\theta\geq 83^o }$) (induced by GZK UHE neutrino) may reach their maximum output  and
 their event maybe observed ; therefore only a
 small fraction ($\sim 30\%$ corresponding to $\sim 50$ $km^3$ mass-water volume for EUSO observation)
 of vertical downward UHE neutrino may be seen by EUSO. This
 signal may be somehow hidden or masked by the more common down-ward UHECR
 showers.  The key reading  signature will be the shower height
  origination: $(\geq 40 km)$ for most downward-horizontal UHECR, $(\leq 10 km)$ for most inclined-horizontal Air UHE
 $\nu$ Induced Shower. A corresponding smaller fraction ($\sim 7.45\%$)  of totally Horizontal
  UHE neutrino Air shower, orphan of their final Cherenkov flash, in competition
  with the horizontal UHECR, may be also clearly observed:
  their observable mass is only $V_{Air-\nu-Hor}$ $\sim 11.1$ $km^3$ for EUSO
  observation duty-cycle. These masses reflect into a characteristic
  threshold behaviour shown by bounds in Figure $7$.
\subsection{ UHE $\nu_{\tau}-\tau$  Double Bang Shower }
 A more  rare, but spectacular, double $\nu_{\tau}$-$\tau$ bang  in Air (comparable in principle to
 the PeVs expected  "double bang" in water \cite{Learned Pakvasa 1995})
 may be exciting, but difficult to be observed; the EUSO effective calorimeter mass for such Horizontal event is only $10\%$ of the
 UHE $\nu$ Horizontal ones (($\sim 1.1$ $ km^3$)); therefore its event rate is nearly excluded
 needing a too high neutrino fluxes see Fig.$6$; indeed it should be also noted that the EUSO energy threshold
($\geq 3\cdot 10^{19}$eV) imply such a very large  ${\tau}$
Lorents boost distance; such large ${\tau}$ track  exceed (by
more than a factor three) the EUSO disk Area diameter ($\sim
450$km); therefore the expected Double Bang
Air-Horizontal-Induced ${\nu}$ Shower thresholds are suppressed
by a corresponding factor as shown in Figure $6$. More abundant
single event  Air-Induced ${\nu}$  Shower (Vertical or Horizontal)
thresholds are facing different Air volumes and  quite different
visibility as shown and summirized in $Fig 7$. It must be taken
into account an additional factor three (because of three light
neutrino states) in the Air-Induced ${\nu}$  Shower  arrival flux
respect to incoming $\nu_{\tau}$ (and $\bar{\nu_{\tau}}$ ),
making the Air target not totally a negligible calorimeter.
\subsection{ UHE $\nu_{\tau}-\tau$  Air  Single Bang Shower }
There are also a sub-category  of $\nu_{\tau}$ - $\tau$ "double
 bang" due to a first horizontal UHE $\nu_{\tau}$ charged current interaction
 in air  nuclei (the first bang) that is lost from the EUSO view;
 their UHE  secondary $\tau$ fly and decay leading to a Second Air-Induced Horizontal Shower, within the EUSO
 disk area. These  horizontal "Double-Single $\tau$ Air Bang"  Showers
 (or if you like popular terminology, these Air-Earth Skimming neutrinos or just Air-HORTAU event)
  are produced within a very wide Terrestrial Crown Air Area whose radius is exceeding $\sim 600- 800$ km
 surrounding  the EUSO Area of view. However it is easy to show
 that they will just double the  Air-Induced ${\nu}$  Horizontal Shower
 rate due to one unique flavour. Therefore the total Air-Induced Horizontal Shower (for
 all $3$ flavours and the additional $\tau$ decay in flight) are summirized and considered
 in Fig.7. The relevant UHE neutrino signal, as discussed below, are due to the
 Horizontal Tau Air-Showers originated within the (much denser)Earth
 Crust:the called  HORTAUs (or Earth Skimming $\nu_{\tau}$).

\begin{figure}\centering\includegraphics[width=8cm]{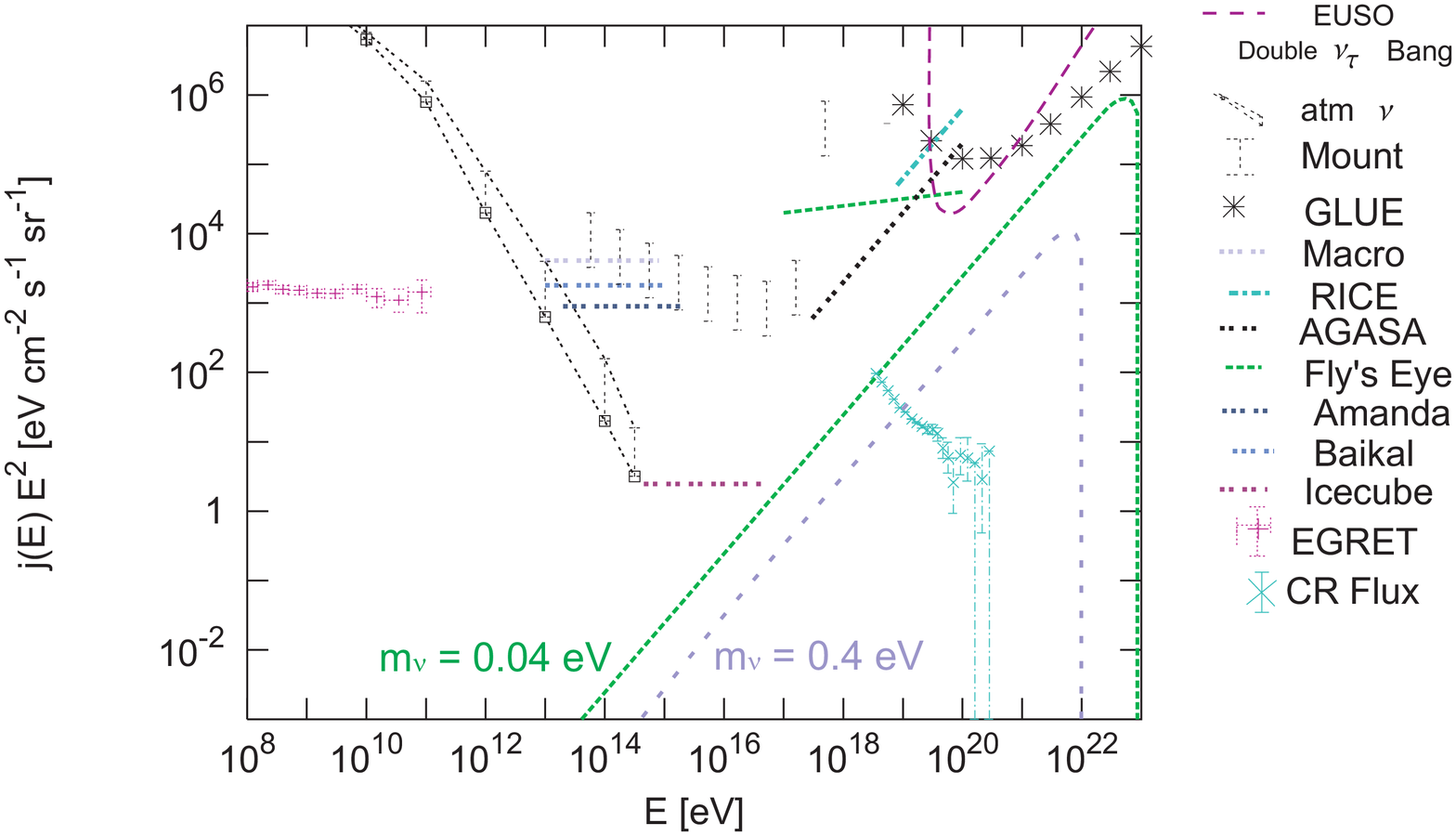}
\caption {EUSO threshold for Double bang $\tau$ Neutrino over
other $\gamma$, $\nu$ and Cosmic Rays (C.R.) Fluence  and bounds
in different energy windows. The  Fluence threshold for EUSO has
been estimated for a three year experiment lifetime. Competitive
experiment are also shown as well as the Z-Shower expected
spectra in most probable light neutrino mass values ($m_{\nu} =
0.04, 0.4$ eV). \cite{Fargion
2000-2002},\cite{Fargion2001a},\cite{Kalashev:2002kx},
\cite{Fargion et all. 2001b},\cite{Fargion 2002d}.}
\label{fig:fig2}
\end{figure}


\begin{figure}\centering\includegraphics[width=8cm]{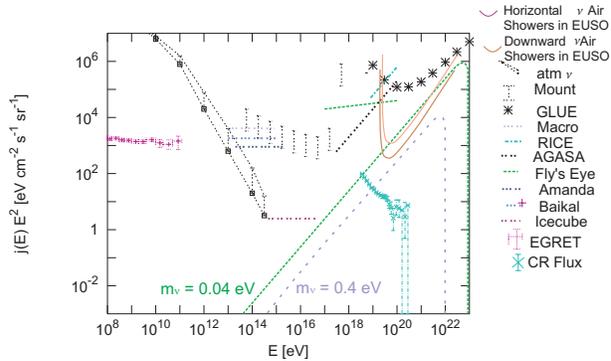}
\caption {EUSO thresholds for Horizontal and Vertical Downward
Neutrino Air induced shower over other $\gamma$, $\nu$ and Cosmic
Rays (C.R.) Fluence  and bounds. The  Fluence threshold for EUSO
has been estimated for a three year experiment lifetime.
Competitive experiments are also shown as well as the Z-Shower
expected spectra in light neutrino mass values ($m_{\nu} = 0.04,
0.4$ eV). \cite{Fargion 2000-2002},
\cite{Fargion2001a},\cite{Kalashev:2002kx}, \cite{Fargion et all.
2001b},\cite{Fargion 2002d}.} \label{fig:fig2}
\end{figure}


\section{UHE $\nu_{\tau}-\tau$ from Earth Skin: HORTAUs}
 As already mention the UHE $\nu$ astronomy maybe greatly amplified by
 $\nu_{\tau}$ appearance via flavour mixing and oscillations. The
  consequent scattering of $\nu_{\tau}$ on the Mountains or into the
 Earth Crust may lead to Horizontal Tau Air-Showers :HORTAUs (or
 so called Earth Skimming Showers \cite{Fargion2001a},\cite{Fargion2001b},\cite{Fargion 2000-2002}\cite{Feng et al 2002}).
 Indeed UHE $\nu_{\tau}$ may skip below
 the Earth and escape as $\tau$ and finally decay in flight, within air atmosphere,
 as well as inside  the Area of  view of EUSO, as shown in
 Figure $8$. Any UHE-GZK Tau Air Shower induced event is approximately born within
a wide  ring  (whose radiuses extend between $R \geq300$ and $R
\leq800$ km from the EUSO Area center). Because of the wide area
and deep $\tau$ penetration \cite{Fargion
2000-2002},\cite{Fargion 2002b},\cite{Fargion 2002d} the amount
of interacting matter where UHE $\nu$ may lead to $\tau$ is huge
($\geq 2 \cdot 10^5$ $km^3$) ;however only a tiny fraction of
these HORTAUs will beam and Shower within the EUSO Area within
EUSO.After carefully estimate (using also results in
\cite{Fargion 2000-2002},\cite{Fargion 2002b}, \cite{Fargion
2002c},\cite{Fargion 2002d}) I probed  a lower bound (in sea
matter) for these effective Volumes respectively at ($1.1 \cdot
10^{19}$eV) and  at ($3 \cdot 10^{19}$ eV) energy:

\begin{equation}
V_{eff}\geq A _{EUSO} \cdot\frac{1}{2} 
\cdot\sin{\delta\tilde{\theta}_{h_{1}}}\cdot
l_{\tau}\cdot\delta\tilde{\theta}_{h_{1}}
\end{equation}
The above geometrical quantities
$\delta\tilde{\theta}_{h_{1}}$,$l_{\tau}$,  are the Earth Skimming
or HORTAU angle and tau interaction lenght defined in reference
\cite{Fargion 2002b} while $A _{EUSO}$ is the EUSO Area. Assuming
a characteristic EUSO radius of $225$ km and at above energies
one obtains a lower bound:
$$V_{eff}\simeq5.13 \cdot 10^{3} km^{3}$$
$$V_{eff}\simeq 6.25 \cdot 10^{3} km^{3}$$
These are the bounds applied in red curves in figures.
   A more  exact and detailed derivation offer a larger Volume:
\begin{equation}
V_{eff}\simeq A _{EUSO} 
\cdot(\sin{\delta\tilde{\theta}_{h_{1}})^{2}\cdot l_{\tau}}
\end{equation}
 These volumes are twice the above considered bounds but are not discussed here.
 Therefore at GZK energies ($1.1 \cdot 10^{19}$eV) the horizontal $\tau$ by HORTAUs are more than $50$
times more abundant than any corresponding Horizontal Air Induced
at energy $10^{19}$eV neutrino air-induced at the same energy.
It should be remind that all these bound for EUSO in figure
are already suppressed by a factor $0.1$ due to minimal EUSO duty cycle.

\begin{figure}\centering\includegraphics[width=8cm]{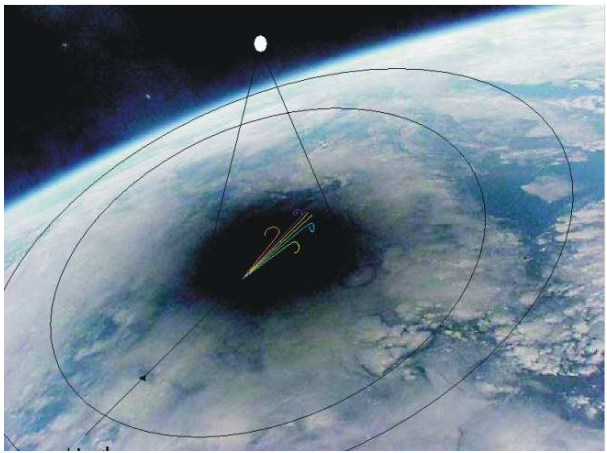}
\caption {A schematic Horizontal High Altitude Shower or similar
Horizontal Tau Air-Shower (HORTAUs) and its open fan-like jets
due to geo-magnetic bending seen from a high quota by EUSO
satellite. The image background is moon eclipse shadow observed
by Mir on Earth. The forked Shower is multi-finger containing a
inner $\gamma$ core and external fork spirals due to $e^+  e^-$
pairs (first opening) and  ${\mu}^+ {\mu}^-$ pairs \cite{Fargion
2000-2002}, \cite{Fargion2001a}, \cite{Fargion2001b}.}
\label{fig:fig2}
\end{figure}

\begin{figure}\centering\includegraphics[width=8cm]{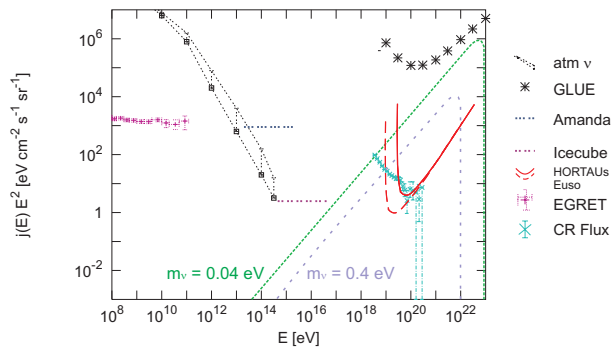}
\caption {EUSO thresholds for Horizontal Tau Air-Shower HORTAUs
(or Earth Skimming Showers) over few $\gamma$, $\nu$ and Cosmic
Rays (C.R.) Fluence and bounds. Dash curves for HORTAUs  are
drawn assuming an EUSO threshold at $10^{19}$eV. Because the
bounded $\tau$ flight distance (due to the contained terrestrial
atmosphere height) the main signal is  better observable at $1.1
\cdot 10^{19}$eV than higher energies. The Fluence threshold for
EUSO has been estimated for a three year experiment lifetime.
Z-Shower or Z-Burst expected spectra in light neutrino mass
values ($m_{\nu} = 0.04, 0.4$ eV) are shown. \cite{Fargion
2000-2002}, \cite{Fargion2001a},\cite{Kalashev:2002kx},
\cite{Fargion et all. 2001b},\cite{Fargion 2002d}.}
\label{fig:fig2}
\end{figure}


\begin{figure}\centering\includegraphics[width=8cm]{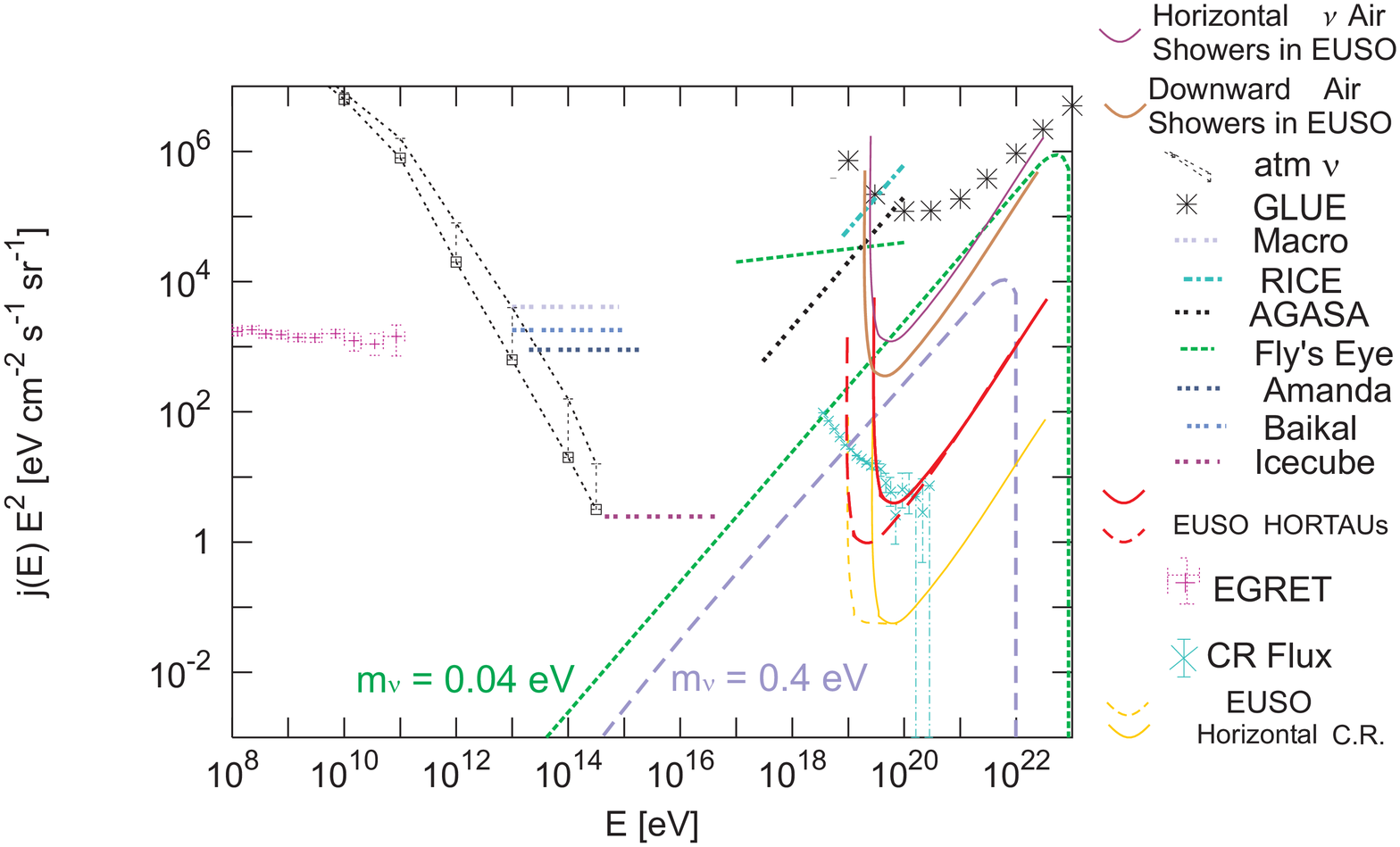}
\caption {EUSO thresholds for Horizontal Tau Air-Shower shower,
HORTAUs (or Earth Skimming Showers) over all other $\gamma$, $\nu$
and Cosmic Rays (C.R.) Fluence and bounds. The Fluence threshold
for EUSO has been estimated for a three year experiment lifetime.
Competitive experiment are also shown as well as the Z-Shower
expected spectra in light neutrino mass values ($m_{\nu} = 0.04,
0.4$ eV). As above dash curves for both HORTAUs and Horizontal
Cosmic Rays are drawn assuming an EUSO threshold at $10^{19}$eV.
\cite{Fargion 2000-2002},
\cite{Fargion2001a},\cite{Kalashev:2002kx}, \cite{Fargion et all.
2001b},\cite{Fargion 2002d}.} \label{fig:fig2}
\end{figure}

However the Air-Shower induced neutrino may reflect all three
light neutrino flavours, while HORTAUs are made only by
$\nu_{\tau}$,$\bar{\nu_{\tau}}$ flavour. Nevertheless the
dominant role of HORTAUs overcome (by a factor $\geq 15$) all
other Horizontal EUSO neutrino event: their  expected event rate
are, at $\Phi_{\nu}\geq 3 \cdot 10^{3}$ eV $cm^{-2} s^{-1}$
neutrino fluence (as in Z-Shower model in Figure $9-10$), a few
hundred event a year and they may already be comparable or even
may exceed the expected Horizontal CR rate. Dash curves for both
HORTAUs and Horizontal Cosmic Rays are drawn assuming an EUSO
threshold at $10^{19}$eV. Because the bounded $\tau$ flight
distance (due to the contained terrestrial atmosphere height) the
main signal is  better observable at $1.1 \cdot 10^{19}$eV than
higher energies as emphasized  in Fig.$9-10$ at different
threshold curves.

\section{Conclusions}

Highest Energy Neutrino signals may be well observable by next
generation satellite as EUSO: the main source of such neutrino
traces are UPTAUs (Upward Tau blazing the telescope born in Earth
Crust) and mainly HORTAUs (Horizontal Tau Air-Showers originated
by an Earth-Skimming UHE $\nu_{\tau}$). These showers will be
opened in a characteristic thin fan-jet ovals like the $8$-shape
horizontal cosmic ray observed on Earth. The UPTAUs will arise
mainly at PeV energies (because the Earth neutrino opacity at
higher energies and because the shorter $\tau$ boosted lenght ,
at lower energies)\cite{Fargion 2000-2002}; UPTAUs will be
detected as a thin stretched multi-pixel event by EUSO, whose
orientation is polarized orthogonal to local geo-magnetic field.
The EUSO sensibility (effective volume ($V_{eff}$$\sim 0.1 km^3$)
for 3 years of detection) will be deeper an order of magnitude
below present AMANDA-Baikal bounds. Horizontal Tau Air-Shower at
GZK energies will be better searched and revealed. They are
originated along huge Volumes around the EUSO Area. Their
horizontal skimming secondary $\tau$ decay occur far away $\geq
500$ km, at high altitude ($\geq 20-40$ km) and it will give
clear signals distinguished from downward horizontal UHECR.
HORTAUs are grown by UHE neutrino interactions inside huge
volumes ($V_{eff}$$\geq 5130-6250 km^3$) respectively for
incoming neutrino energy $E_{\nu_{\tau}}$ $\simeq 10^{19}$ eV and
$3 \cdot10^{19}$ eV. To obtain these results we applied the
procedure described in recent articles \cite{Fargion
2002b},\cite{Fargion 2002c},\cite{Fargion 2002d}. As summirized
in last Figures the expected UHE fluence $$\Phi_{\nu}\simeq
10^{3} eV cm^{-2} s^{-1}$$ needed in most Z-Shower models (as
well as in most topological relic scenario) to solve GZK puzzles,
will lead to nearly a hundred of horizontal events a year
comparable to UHECR ones. Even in the most conservative scenario
where a minimal GZK-$\nu$ fluence must take place (at least at
$$\Phi_{\nu} \simeq 10 eV cm^{-2} s^{-1}$$ ,just comparable to
well observed Cosmic Ray fluence), a few or a ten of such UHE
astrophysical neutrino must be observed (respectively at
$10^{19}$ eV and $3 \cdot10^{19}$ eV energy windows) during three
year of EUSO data recording. To improve their visibility EUSO
must, in
our opinion one may:\\
 a) Improve the fast pattern recognition of Horizontal Shower
Tracks with their few distant dots with forking signature.\\
 b) Enlarge the Telescope
Radius to embrace also lower $10^{19}$ eV energy thresholds where UHE neutrino signals are enhanced.\\
 c) Consider a detection  at  angular $\Delta\theta$ and at height $\Delta h$ level within an
accuracy $\Delta\theta \leq 0.2^o$,$\Delta h \leq 2$ km.\\ Even
all  the above results have been derived carefully  following
\cite{Fargion 2002b},\cite{Fargion 2002c},\cite{Fargion 2002d} in
a minimal realistic framework they may be used within $10\%$
nominal value due to the present uncertain in  EUSO detection
capabilities.



\subsection*{Acknowledgment}
The author wish to thank Prof. Livio Scarsi  for inspiring the
present search  as well as the EUSO collaboration for the exciting
discussion during  November workshop in Rome; the author thanks
also C.Leto and P.G.De Sanctis Lucentini and M.Teshima for support
and technical suggestions.

%
\bibliography{xbib}
\end{document}